\documentclass[fleqn,twoside]{article}
\usepackage[headings]{espcrc2}

\readRCS
$Id: espcrc2.tex,v 1.2 2004/02/24 11:22:11 spepping Exp $
\usepackage{graphicx}
\usepackage[figuresright]{rotating}

\def\mathswitch#1{\relax\ifmmode#1\else$#1$\fi}
\def\mathswitchr#1{\relax\ifmmode{\mathrm{#1}}\else$\mathrm{#1}$\fi}

\newcommand{\als}{\alpha_{\mathrm{s}}}

\newcommand{\GF}{\mathswitch {G_\mu}}
\newcommand{\sw}{\mathswitch {s_\rw}}
\newcommand{\rT}{{\mathrm{T}}}

\newcommand{\rw}{\mathrm{w}}

\newcommand{\de}{\delta}

\newcommand{\PH}{\mathswitchr H}
\newcommand{\PW}{\mathswitchr W}
\newcommand{\PZ}{\mathswitchr Z}

\def\mathswitch#1{\relax\ifmmode#1\else$#1$\fi}
\newcommand{\MH}{\mathswitch {M_\PH}}
\newcommand{\MW}{\mathswitch {M_\PW}}
\newcommand{\MZ}{\mathswitch {M_\PZ}}

\newcommand{\GeV}{\unskip\,\mathrm{GeV}}

\def\reffi#1{\mbox{Figure~\ref{#1}}}

\def\refta#1{\mbox{Table~\ref{#1}}}
\def\citere#1{\mbox{Ref.~\cite{#1}}}
\def\citeres#1{\mbox{Refs.~\cite{#1}}}

\newcommand{\lsim}
{\;\raisebox{-.3em}{$\stackrel{\displaystyle <}{\sim}$}\;}

\newcommand{\hhtwoloops}{\GF^2\MH^4}

\newcommand{\AmS}{{\protect\the\textfont2
  A\kern-.1667em\lower.5ex\hbox{M}\kern-.125emS}}

\title{Strong and electroweak NLO corrections to
Higgs-boson production in vector-boson fusion at the LHC}

\author{Mariano Ciccolini\address[PSI]{Paul Scherrer Institut,
    W\"urenlingen und Villigen, CH-5232 Villigen PSI, Switzerland},
  Ansgar Denner\addressmark[PSI] and Stefan
  Dittmaier\address{Max-Planck-Institut f\"ur Physik
    (Werner-Heisenberg-Institut), D-80805 M\"unchen, Germany}\thanks{
    This work is supported in part by the European Community's
    Marie-Curie Research Training Network HEPTOOLS under contract
    MRTN-CT-2006-035505.}}
       
\runtitle{NLO corrections to Higgs-boson production in vector-boson
  fusion at the LHC} 
\runauthor{M. Ciccolini, A. Denner and S. Dittmaier}

\begin{document}

\begin{abstract}
  We present results on the strong and electroweak NLO corrections to
  the production of a Higgs boson plus two hard jets via weak
  interactions at the LHC.  The calculation includes all weak-boson
  fusion and quark--antiquark annihilation diagrams as well as all
  related interferences.  We discuss corrections of different origin
  (QCD corrections of vector-boson-fusion type and interferences,
  electroweak corrections induced by quark or photonic initial states,
  heavy-Higgs-boson effects, etc.)  and give some new results for
  distributions for a Higgs-boson mass of $200\GeV$.  The electroweak
  corrections
  are of the same size as the QCD corrections, viz.\ typically at the
  level of $5{-}10\%$ for a Higgs-boson mass up to $\sim700\GeV$.  In
  general, they do not simply rescale differential distributions, but
  induce distortions at the level of 10\%.  The discussed corrections
  have been implemented in a flexible Monte Carlo event generator.
  \vspace{1pc}
\end{abstract}

\maketitle

\section{Introduction}

The electroweak (EW) production of a Standard Model Higgs boson in
association with two hard jets in the forward and backward regions of
the detector---frequently quoted as ``vector-boson fusion'' (VBF)---is
a cornerstone in the Higgs-boson search both in the ATLAS \cite{Asai:2004ws}
and CMS \cite{Abdullin:2005yn} experiments at the LHC and also plays
an important role in the determination of Higgs-boson couplings at this
collider.

Higgs+2jets production in pp collisions proceeds through two different
channels.  The first channel corresponds to a pure EW process. It
comprises the scattering of two (anti-)quarks mediated by $t$- and
$u$-channel W- or Z-boson exchange, with the Higgs boson radiated off
the virtual weak boson.  It also involves Higgs-boson radiation off a
W- or Z-boson produced in $s$-channel quark--antiquark annihilation
(Higgs-strahlung process), with the weak boson decaying hadronically.
The second channel proceeds through strong interactions, the Higgs
boson being radiated off a heavy-quark loop that couples to any
\looseness -1 parton of the incoming hadrons via
gluons~\cite{DelDuca:2001fn}.

In the weak-boson-mediated processes, the two scattered quarks are
usually visible as two hard forward jets, in contrast to other jet
production mechanisms, offering a good background suppression
(transverse-momentum and rapidity cuts on jets, jet rapidity gap,
central-jet veto, etc.).  Applying appropriate event selection
criteria (see e.g.\ \citere{Barger:1994zq} and references in
\citeres{Spira:1997dg,Djouadi:2005gi}) it is possible to sufficiently
suppress background and to enhance the VBF channel over the hadronic
Higgs+2jets production mechanism.

In order to match the required precision for theoretical predictions
at the LHC, QCD and EW corrections are needed. When VBF cuts are
imposed, the cross section can be approximated by the contribution of
squared $t$- and $u$-channel diagrams only, which reduces the QCD
corrections to vertex corrections to the weak-boson--quark coupling.
Explicit next-to-leading order (NLO) QCD calculations in this
approximation exist since more than a decade
\cite{Spira:1997dg,Han:1992hr}, while corrections to distributions
have been calculated in the last few years
\cite{Figy:2003nv,Figy:2004pt}.  Recently, the full NLO EW and QCD
corrections to this process have become available
\cite{Ciccolini:2007jr,Ciccolini:2007ec}. This calculation includes,
for the first time, the complete set of EW and QCD diagrams, namely
the $t$-, $u$-, and $s$-channel contributions, as well as all
interferences at NLO. Higher-order loop-induced interference effects
between VBF and gluon--gluon fusion have been examined in
\citeres{Andersen:2007mp} and turn out to be completely negligible.
Very recently, also the additional supersymmetric QCD and EW
corrections within the MSSM have been evaluated~\cite{Hollik:2008xn}
and found to be typically below or at the 1\% level.

In these proceedings we briefly summarize the calculation of the NLO
EW and QCD corrections presented in
\citeres{Ciccolini:2007jr,Ciccolini:2007ec} and give new results
on distributions for a Higgs-boson mass of $200\GeV$.

\section{Brief outline of the NLO calculation}

We have calculated the complete QCD and EW NLO corrections to
Higgs-boson
\looseness -1
production via weak VBF at the LHC. At LO, this process receives
contributions from the partonic processes $qq\to\PH qq$,
$q\bar{q}\to\PH q\bar{q}$, and $\bar{q}\bar{q}\to\PH\bar{q}\bar{q}$.
All LO and one-loop NLO diagrams are related by crossing symmetry to
the corresponding decay amplitude $\PH\to q\bar{q}q\bar{q}$. The QCD
and EW NLO corrections to these decays were discussed in
\citere{Bredenstein:2006rh}.  In our calculation of the corrections,
which is described in detail in
\citeres{Ciccolini:2007jr,Ciccolini:2007ec}, we partially made use of
the results on the related Higgs decays.  The electroweak NLO
corrections have been supplemented by the leading two-loop
heavy-Higgs-boson effects~\cite{Ghinculov:1995bz} proportional to
$\GF^2\MH^4$.

In the $s$-channel diagrams intermediate W and Z~bosons can become
resonant, corresponding to $\PW\PH/\PZ\PH$ production with subsequent
gauge-boson decay. In order to consistently include these resonances,
we use the ``complex-mass scheme'' at the one-loop
level~\cite{Denner:2005fg}, which respects all relations that follow
from gauge invariance. In this approach the W- and Z-boson masses are
consistently considered as complex quantities, defined as the
locations of the propagator poles in the complex plane.  The tensor
integrals are evaluated using the reduction techniques of
Refs.~\cite{Denner:2002ii,Denner:2005nn}, which include direct
reductions of pentagon integrals to boxes and specific methods to
treat exceptional phase-space configurations in a numerically stable
way.

Real corrections consist of gluon and photon emission and processes
with $gq$ and $\gamma q$ initial states.  The mass singularities from
collinear initial-state splittings are absorbed via factorization by
the usual PDF redefinition both for the QCD and photonic corrections.
Technically, the soft and collinear singularities are isolated in the
dipole subtraction method following
\citeres{Dittmaier:1999mb,Dittmaier:2008md}, and the result was
checked with the phase-space slicing method.

Each part of the whole calculation has been worked out twice and
independently, and all corrections are implemented in a flexible Monte
Carlo event generator based on multi-channel integration.

\section{Numerical results}

Numerical results for the Higgs-mass dependence and the scale
dependence of the total cross section with and without VBF cuts as
well as distributions for $\MH=120\GeV$ have been presented in
\citeres{Ciccolini:2007jr,Ciccolini:2007ec}. Some distributions for
$\MH=200\GeV$ have been published in \citere{Ciccolini:2007tq}. Here
we summarize the size of the different contributions to the total
cross section and show some more distributions for $\MH=200\GeV$.

All presented results are based on the input parameters as given in
\citere{Ciccolini:2007ec}.  Since quark-mixing effects are suppressed,
the CKM matrix is set to the unit matrix.  The electromagnetic
coupling is fixed in the $G_\mu$ scheme, i.e.\ it is set to
$\alpha_{\GF}=\sqrt{2}\GF\/\MW^2\sw^2/\pi$, because this accounts for
electromagnetic running effects and some universal corrections of the
$\rho$ parameter.
We use the MRST2004QED PDF \cite{Martin:2004dh} which consistently
include ${\cal O}(\alpha)$ QED corrections and a photon distribution
function for the proton. In our default set-up we use only four quark
flavours for the external partons, i.e.\ we do not take into account
the contribution of bottom quarks, which is suppressed.  Partonic
processes involving b~quarks are, however, included in our code in LO.
The renormalization and factorization scales are set to $\MW$, 5
flavours are included in the two-loop running of $\als$, and
$\als(\MZ)=0.1187$.  We apply typical VBF cuts to the outgoing jets as
described in detail in \citere{Ciccolini:2007ec}.
 
In \refta{ta:xsection-contr} we summarize the impact of different
contributions to the integrated cross sections with and without VBF
cuts for Higgs masses between 120 and $200\GeV$, for $\MH=400\GeV$,
and for $\MH=700\GeV$ in per cent of the corresponding LO cross
section.

Previous calculations of the VBF process
\cite{Spira:1997dg,Han:1992hr,Figy:2003nv,Figy:2004pt} have
consistently neglected $s$-channel contributions (``Higgs
strahlung''), which involve diagrams where one of the vector bosons
can become resonant, as well as the interference between $t$- and
$u$-channel fusion diagrams.  For small Higgs-boson masses the
contributions of $s$-channel diagrams, $\Delta_{s-\mathrm{channel}}$,
range between 10\% and 30\% when no cuts are applied.  With VBF cuts,
the $s$-channel contributions are strongly suppressed, yielding less
than $0.6\%$ of the cross section for all the studied Higgs-boson
masses. The contributions from interferences between $t$- and
$u$-channel diagrams, $\Delta_{t/u-\mathrm{int}}$, are below $1\%$
with or without VBF cuts and thus negligible.  Consequently, applying
typical experimental VBF cuts, the contributions from $s$-channel
diagrams and $t/u$-channel interferences can be safely neglected.

Next, we list in \refta{ta:xsection-contr} the contributions arising
at LO from processes that include b-quarks in the initial and/or final
states. These are included in our programs but not in the default
set-up. For Higgs-boson masses below $ 200\GeV$, they increase the
total cross section without cuts by about 4\% and the one with cuts by
about 2\%. For larger Higgs-boson masses their impact is smaller.

In the lower part of \refta{ta:xsection-contr} contributions of
various NLO corrections are listed. Both electroweak and QCD
corrections are at the level of 5--10\%.  The QCD corrections are
dominated by the previously known diagonal contributions, i.e.\ by the
vector-boson--quark--antiquark vertex corrections to squared LO
diagrams, $\de_{\mathrm{QCD(diag)}}$. All other QCD contributions,
i.e.\ QCD corrections to interferences between the different LO
diagrams and interferences with gg-fusion and g-splitting diagrams
(see \citere{Ciccolini:2007ec} for a precise definition), summarized
in $\de_{\mathrm{QCD(int)}}$, are at the per-mille level and even
partially cancel each other.  They are not enhanced by contributions
of two $t$- or $u$-channel vector bosons with small virtuality and
therefore even further suppressed when applying VBF cuts.  For the
electroweak corrections we give the impact of the quark--quark induced
processes, $\de_{\mathrm{EW,}qq}$, and the one of the photon induced
processes, $\delta_{\mathrm{EW,} q\gamma}$ separately. The latter turn
out to be $\sim+1\%$ and reduce the complete electroweak corrections
for small Higgs-boson masses. The dominant two-loop correction
\smash{$\de_{\hhtwoloops}$} due to Higgs-boson self-interaction, which
is contained in $\de_{\mathrm{EW,}qq}$, is completely negligible in
the low-$\MH$ region, but becomes important for large Higgs-boson
masses and yields $+4\%$ for $\MH=700\GeV$ which constitutes about
$50\%$ of the total EW corrections.  Obviously for Higgs masses in
this region and above the perturbative expansion breaks down, and the
two-loop factor $\de_{\hhtwoloops}$ might serve as an estimate of the
theoretical uncertainty.
\begin{table*}
\caption{Impact of specific corrections to the
  cross section without and with VBF cuts relative to LO.}
\label{ta:xsection-contr}
\renewcommand{\tabcolsep}{1pc} 
\renewcommand{\arraystretch}{1.1} 
\begin{center}
\begin{tabular}{rrrrrrr}
\hline
& \multicolumn{3}{c}{no cuts} & \multicolumn{3}{c}{VBF cuts}
\\
$\MH[\GeV]$ & $120{-}200$ & $400$ & $700$ & 
\quad $120{-}200$ & $400$ & $700$
\\
\hline
$\Delta_{s-\mathrm{channel}}[\%]$ & $30{-}10$ & $2$ & $1$ 
& $<0.6$ & $<0.3$ & $<0.1$
\\
$\Delta_{t/u-\mathrm{int}}[\%]$ & $<0.5$ & $<0.1$ & $<0.1$ & $<0.1$ & 
$<0.1$ & $<0.1$ 
\\
\hline
$\Delta_{\mathrm{b-quarks}}[\%]$ & $\approx4$ & $2$ & $1$ & 
$\approx2$ & $2$ & $1$ 
\\
\hline
$\de_{\mathrm{QCD(diag)}}[\%]$ & $4{-}0.5$ & $-0$ & $+1$ & $\approx-5$ 
& $-6$ & $-7$
\\
$\de_{\mathrm{QCD(int)}}[\%]$ & $\lsim0.2$ & $-0.2$& $-0.1$ 
& $<0.1$ & $<0.1$ & 
$<0.1$
\\
$\de_{\mathrm{EW,}qq}[\%]$ & $\approx-5$ & $-5$ &  $+6$ &
$\approx-7$ & $-5$ & $+5$
\\
$\de_{\mathrm{EW,}q\gamma}[\%]$ & $\approx+1$ & $+2$ & $+2$ 
& $\approx+1$ & $+1$ & $+2$
\\
$\de_{\GF^2\MH^4}[\%]$ & $<0.1$ & $+0.4$ & $+4$ & $<0.1$ & $+0.4$ & 
$+4$
\\
\hline
\end{tabular}
\end{center}
\end{table*}

The EW corrections to distributions for $\MH=200\GeV$ are
qualitatively similar to those for $\MH=120\GeV$ presented in
\citere{Ciccolini:2007ec}.  The distributions in the transverse
momentum $p_{\mathrm{j}_1,\rT}$ of the harder tagging jet
$\mathrm{j}_1$ (jet with highest $p_{\rT}$ passing all cuts) and the
distribution in the azimuthal angle separation of the two tagging jets
for $\MH=200\GeV$ have been presented in \citere{Ciccolini:2007tq}.
Here we show some additional distributions for $\MH=200\GeV$.

\begin{figure*}
  \includegraphics[bb= 85 440 285 660, scale=1]{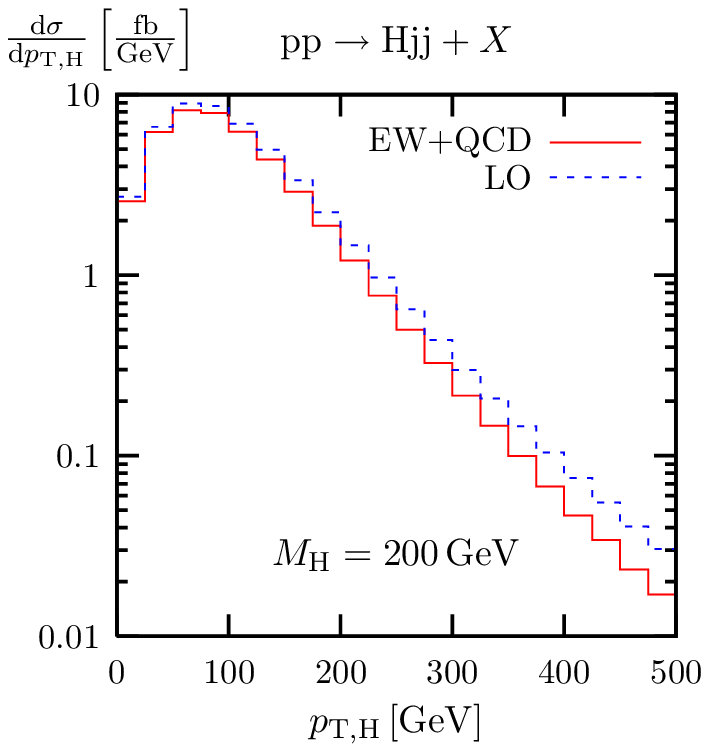} \quad
  \includegraphics[bb= 85 440 285 660, scale=1]{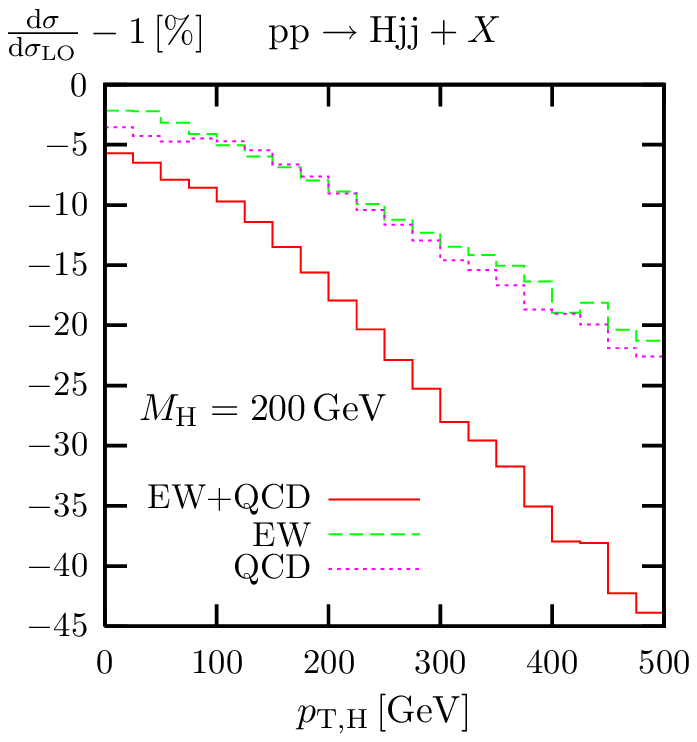}
\vspace*{-3em}
\caption{Distribution in the transverse momentum $p_{\rT,\PH}$ of the
  Higgs boson (left) and corresponding relative corrections (right)
  for $\MH=200\GeV$.}
\label{fi:pth}
\end{figure*}

\begin{figure*}
\includegraphics[bb= 85 440 285 660, scale=1]{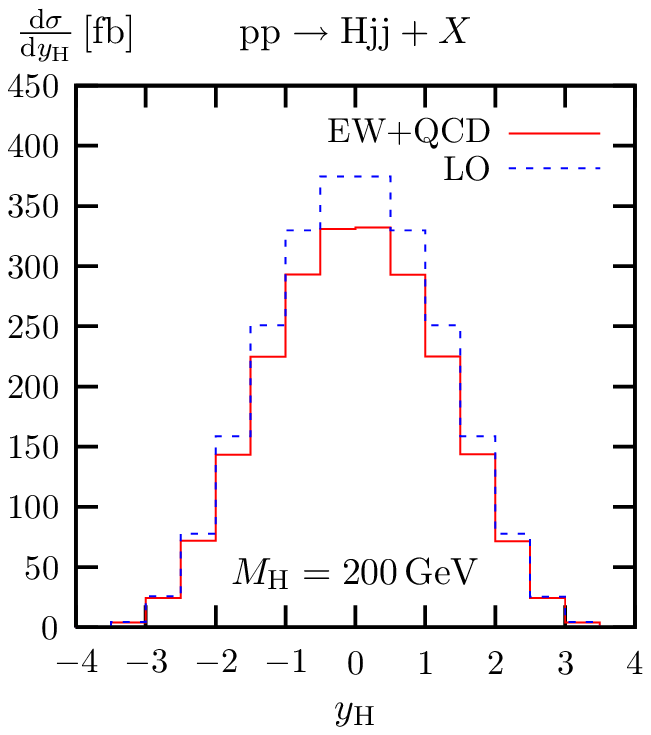}
\quad
\includegraphics[bb= 85 440 285 660, scale=1]{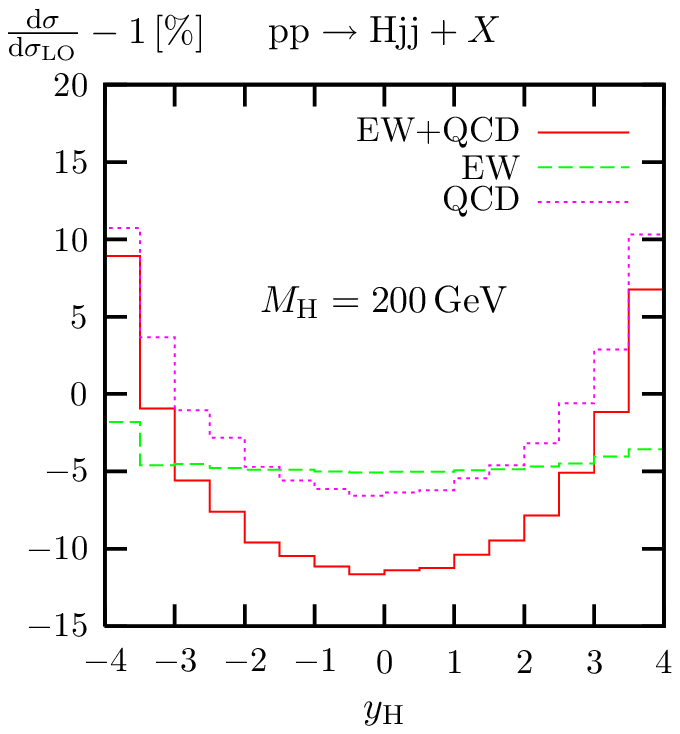}
\vspace*{-3em}
\caption{Distribution in the rapidity $y_{\PH}$ of the Higgs boson (left)
  and corresponding relative corrections (right) for $\MH=200\GeV$.}
\label{fi:mwp}
\end{figure*}

\begin{figure*}
  \includegraphics[bb= 85 440 285 660, scale=1]{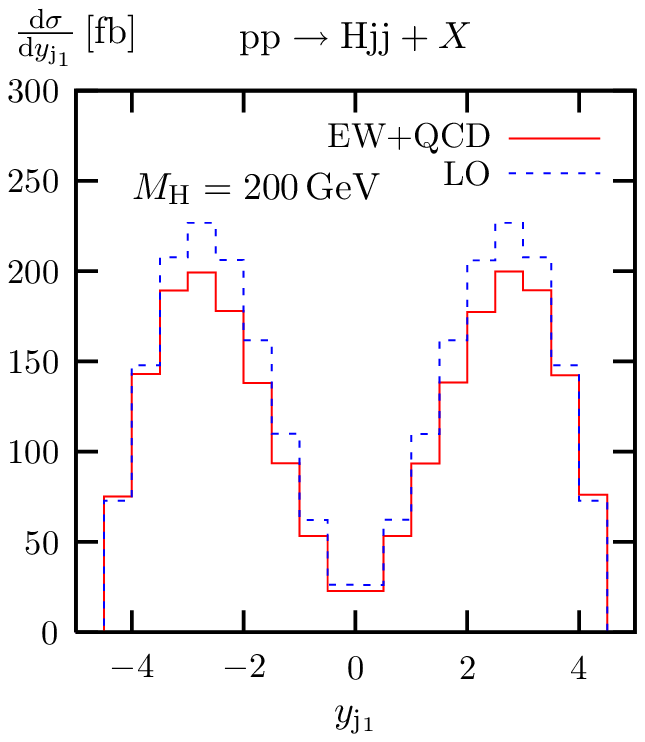}
  \quad \includegraphics[bb= 85 440 285 660, scale=1]{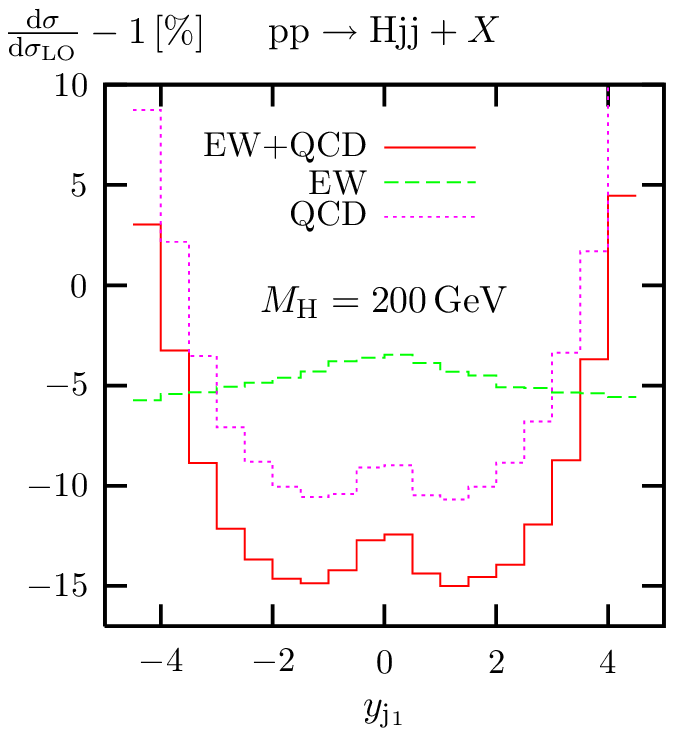}
\vspace*{-3em}
\caption{Distribution in the rapidity $y_{\mathrm{j_1}}$ of the
  harder tagging jet (left) and corresponding relative corrections
  (right) for $\MH=200\GeV$.}
\label{fi:y1}
\end{figure*}

In \reffi{fi:pth} we provide the distribution in the transverse
momentum $p_{\rT,\PH}$ of the Higgs boson. 
The differential cross
section drops strongly with increasing $p_{\rT,\PH}$. As for
$\MH=120\GeV$ both the relative EW and QCD corrections increase in
size and reach $-20\%$ for $p_{\rT,\PH}=500\GeV$.
Figure~\ref{fi:mwp} shows the rapidity distribution of the
Higgs boson.
While the QCD corrections distort this shape by about 10\%, the relative
EW corrections turn out to to be flat where the distribution is
sizeable.
In \reffi{fi:y1}, we depict the distribution in the rapidity of the
harder tagging jet.  
It can be clearly seen that the tagging jets are forward and backward
located.  The EW corrections vary between $-3\%$ and $-6\%$.  The QCD
corrections exhibit a strong dependence on the jet rapidity. They are
about $-10\%$ in the central region but become positive for large
rapidities, where they tend to compensate the EW corrections.  Shape
changes due to the full corrections reach $10\%$.

\section{Conclusions}

Radiative corrections of strong and electroweak interactions have been
discussed at next-to-leading order for Higgs-boson production via
vector-boson fusion at the LHC. All discussed effects have been
implemented into a flexible Monte Carlo generator.  The electroweak
corrections affect the cross section by $5\%$, and are thus as
important as the QCD corrections in this channel. They do not simply
rescale distributions but induce distortions at the level of 10\%.
Effects from photon-induced processes, $s$-channel contributions, and
interferences are small once vector-boson fusion cuts are applied. For
intermediate Higgs-boson masses the remaining theoretical uncertainty
is below the uncertainty from PDFs and expected experimental errors.

\end{document}